\documentstyle[epsfig]{mn}
\newif\ifAMStwofonts

\ifoldfss
  \ifCUPmtlplainloaded \else
    \NewTextAlphabet{textbfit} {cmbxti10} {}
    \NewTextAlphabet{textbfss} {cmssbx10} {}
    \NewMathAlphabet{mathbfit} {cmbxti10} {} 
    \NewMathAlphabet{mathbfss} {cmssbx10} {} 
  \fi
  \ifAMStwofonts
    \ifCUPmtlplainloaded \else
      \NewSymbolFont{upmath} {eurm10}
      \NewSymbolFont{AMSa} {msam10}
      \NewMathSymbol{\upi}     {0}{upmath}{19}
      \NewMathSymbol{\umu}     {0}{upmath}{16}
      \NewMathSymbol{\upartial}{0}{upmath}{40}
      \NewMathSymbol{\leqslant}{3}{AMSa}{36}
      \NewMathSymbol{\geqslant}{3}{AMSa}{3E}

      \let\leq=\leqslant 
       
    \fi
  \fi
\fi 

\ifnfssone
  \newmathalphabet{\mathit}
  \addtoversion{normal}{\mathit}{cmr}{m}{it}
  \addtoversion{bold}{\mathit}{cmr}{bx}{it}
  \newmathalphabet{\mathbfit} 
  \addtoversion{normal}{\mathbfit}{cmr}{bx}{it}
  \addtoversion{bold}{\mathbfit}{cmr}{bx}{it}
  \newmathalphabet{\mathbfss} 
  \addtoversion{normal}{\mathbfss}{cmss}{bx}{n}
  \addtoversion{bold}{\mathbfss}{cmss}{bx}{n}
  \ifAMStwofonts
    \ifCUPmtlplainloaded \else
      \UseAMStwoboldmath
      \makeatletter
      \new@mathgroup\upmath@group
      \define@mathgroup\mv@normal\upmath@group{eur}{m}{n}
      \define@mathgroup\mv@bold\upmath@group{eur}{b}{n}
      \edef\UPM{\hexnumber\upmath@group}
      \new@mathgroup\amsa@group
      \define@mathgroup\mv@normal\amsa@group{msa}{m}{n}
      \define@mathgroup\mv@bold\amsa@group{msa}{m}{n}
      \edef\AMSa{\hexnumber\amsa@group}
      \makeatother
      \mathchardef\upi="0\UPM19
      \mathchardef\umu="0\UPM16
      \mathchardef\upartial="0\UPM40
      \mathchardef\leqslant="3\AMSa36
      \mathchardef\geqslant="3\AMSa3E

      \let\leq=\leqslant 

    \fi
  \fi
\fi 

\ifnfsstwo
  \DeclareMathAlphabet{\mathbfit}{OT1}{cmr}{bx}{it}
  \SetMathAlphabet\mathbfit{bold}{OT1}{cmr}{bx}{it}
  \DeclareMathAlphabet{\mathbfss}{OT1}{cmss}{bx}{n}
  \SetMathAlphabet\mathbfss{bold}{OT1}{cmss}{bx}{n}
  \ifAMStwofonts
    \ifCUPmtlplainloaded \else
      \DeclareSymbolFont{UPM}{U}{eur}{m}{n}
      \SetSymbolFont{UPM}{bold}{U}{eur}{b}{n}
      \DeclareSymbolFont{AMSa}{U}{msa}{m}{n}
      \DeclareMathSymbol{\upi}{0}{UPM}{"19}
      \DeclareMathSymbol{\umu}{0}{UPM}{"16}
      \DeclareMathSymbol{\upartial}{0}{UPM}{"40}
      \DeclareMathSymbol{\leqslant}{3}{AMSa}{"36}
      \DeclareMathSymbol{\geqslant}{3}{AMSa}{"3E}

      \let\leq=\leqslant 

    \fi
  \fi
\fi 

\ifCUPmtlplainloaded \else
  \ifAMStwofonts \else 
    \def\upi{\pi}
    \def\umu{\mu}
    \def\upartial{\partial}
  \fi
\fi

\title{Constraining the Star Formation Histories of Spiral Bulges.}
\author[R. Proctor et al.]
  {R.N. Proctor, $^1$ A.E. Sansom,$^1$ and I.N. Reid.$^2$\\
   $^1$ Centre for Astrophysics, University of Central Lancashire,
   Preston, PR1 2HE, UK\\
   $^2$ California Institute of Technology , 105-24, Pasadena, CA
   91126,USA}

\date{}
\pagerange{\pageref{firstpage}--\pageref{lastpage}}
\pubyear{}

\def\LaTeX{L\kern-.36em\raise.3ex\hbox{a}\kern-.15em
    T\kern-.1667em\lower.7ex\hbox{E}\kern-.125emX}

\begin{document}

\label{firstpage}

\maketitle

\begin{abstract}
Stellar populations in spiral  bulges are investigated using  the Lick
system of spectral   indices. Long-slit spectroscopic  observations of
line-strengths and kinematics made along the minor axes of four spiral
bulges  are   reported.   Comparisons   are    made  between   central
line-strengths in spiral bulges and those in other morphological types
(elliptical, spheroidal (Sph) and   S0). The bulges  investigated  are
found to have central  line-strengths comparable with those  of single
stellar     populations of    approximately    solar   abundance    or
above. Negative  radial   gradients are  observed  in  line-strengths,
similar to those exhibited by elliptical galaxies.  The bulge data are
also consistent  with correlations between Mg$_{2}$, Mg$_{2}$ gradient
and central velocity dispersion   observed in elliptical galaxies.  In
contrast to elliptical galaxies, central line-strengths lie within the
loci defining the range of $<$Fe$>$ and Mg$_{2}$ achieved by Worthey's
\shortcite{W94}  solar   abundance ratio,  single  stellar populations
(SSPs).  The    implication of    solar   abundance   ratios indicates
significant   differences in  the  star formation  histories of spiral
bulges and elliptical galaxies.  A ``single zone with in-fall''  model
of galactic  chemical evolution, using Worthey's \shortcite{W94} SSPs,
is used  to constrain the  possible star  formation histories  of  our
sample.  We   show   that   the $<$Fe$>$,  Mg$_{2}$     and H${\beta}$
line-strengths  observed in  these  bulges cannot  be reproduced using
primordial collapse models  of formation but \emph{can}  be reproduced
by models  with extended in-fall of gas  and star formation (2-17 Gyr)
in   the region  modelled.  One  galaxy  (NGC 5689)   shows   a central
population with  a  luminosity weighted average  age of  $\sim$ 5 Gyr,
supporting  the   idea  of    extended   star formation.     Kinematic
substructure, possibly associated with a central spike in metallicity,
is observed at the centre of the Sa galaxy NGC 3623.
\end{abstract}


\section{Introduction}
\label{intro}
The  recently    developed  Lick     system  of    spectral    indices
\cite{FFBG85,Wea94} has provided powerful  tools for the investigation
of  composite  populations.   Diagnostic plots,   such  as $<$Fe$>$ vs
H${\beta}$ and Mg$_{2}$ vs $<$Fe$>$  , (where $<$Fe$>$ is the  average
of Fe5270 and Fe5335 indices) break the age/metallicity degeneracy and
have illustrated  characteristics  and trends  in elliptical  galaxies
that   constrain their possible  star formation  histories (SFH). This
paper compares  the  line-strengths in four  spiral  bulges with those
observed in other morphologies and places constraints on possible SFHs
in the  sample. Comparison  of  the SFH  of  spiral bulges to  that of
elliptical galaxies is   informative   as the  two  morphologies  show
numerous   similarities.   Bulges exhibit  similar  surface brightness
profiles  and kinematic structures    to   ellipticals and  form    an
overlapping continuation of   the elliptical locus  in the Fundamental
Plane \cite{BBF93}. Colours and  colour gradients in  the two types of
spheroid are  also similar  \cite{BP94}. However, despite similarities
in physical and photometric properties, the two galaxy types are found
in very  different   environments, with the  majority  of  ellipticals
located in clusters, while spiral galaxies are preferentially found in
the field. Another  important   factor in the environment    of spiral
bulges  is the presence  of the  disc,  which severely limits possible
merger histories.\\

Observations of elliptical galaxies   have shown them to possess  high
central   line-strengths.   Gradients are   observed   in  metallicity
sensitive features indicative  of     dissipation in the     formation
process.   The  H${\beta}$ index   shows  no  indication of systematic
gradients in age  (Carollo, Danziger \& Buson  1993; Davies, Sadler \&
Peletier 1993; Fisher, Franx \&  Illingworth 1996; Gorgas et al. 1997;
Vazdekis et   al.  1997).  Modelling  of  Lick  indices  in elliptical
galaxies, for comparison  to observation, has  been  carried out  by a
number  of  authors  \cite{WDJ96,G97,Vea97,SP98a}. These  studies have
shown that  the high line-strengths found at  the centre of elliptical
galaxies can only be  achieved if the bulk  of the observed population
forms from pre-enriched gas. Two possible processes have been proposed
to produce this  pre-enrichment.  One possibility  is an initial  mass
function (IMF) biased  towards high mass stars in  the early stages of
galaxy  formation. In  this way  the   first few generations  of stars
generate large  quantities of metals \cite{GM97,Vea97}. Alternatively,
a period of star formation  (SF), and  thus interstellar medium  (ISM)
enrichment, with  a Salpeter IMF prior  to the main burst, can produce
the necessary pre-enrichment.  This delayed burst of SF may correspond
to   a   merger/interaction   model of   elliptical   galaxy formation
\cite{SP98a}.

The second important feature of line-strengths in ellipticals is their
high central Mg$_{2}$ with respect to $<$Fe$>$ when compared to models
assuming   solar abundance  ratios \cite{WFG92,DSP93,Gea97,G97}.  This
Mg$_{2}$  excess is interpreted   as  indicating [Mg/Fe] greater  than
solar  \cite{DSP93}.  High [Mg/Fe] can be  achieved  by star formation
times of $\la$ 1 Gyr, in accordance with the  short time-scales of gas
inflow in  both merger/interaction \cite{BH96} and primordial collapse
models  of elliptical galaxy   formation  (Theis, Burkert \&  Hensler,
1992). Short bursts of star formation ensure that the metal enrichment
of the ISM  from  which the bulk of   the population forms,  is mainly
produced   in the  Type  II  supernovae that  are  the main  sites for
production of  Mg. The biased  IMF proposal also naturally produces an
excess of  Type II supernovae (and  consequently  Mg$_{2}$) due  to an
over-production  of high mass stars. While  both merger and primordial
collapse (with    biased IMF) models   can  successfully reproduce the
indices in the Mg$_{2}$ vs $<$Fe$>$ plane, the high central H${\beta}$
values of   many  elliptical galaxies are   a challenge  to primordial
collapse  models.  In  contrast,  merger  models  clearly  predict the
existence of younger populations in these objects.

Observations  of   other    galaxy    morphologies, such     as    S0s
\cite{S93,FFI96}   and  Sphs  \cite{Gea97}, show  differences  between
central  line-strengths  in various galaxy types   in  the $<$Fe$>$ vs
Mg$_{2}$ plane.   S0s tend  to   exhibit weaker Mg$_{2}$  for a  given
$<$Fe$>$ than do  Es  while still showing  an  excess with respect  to
values for solar abundance ratio SSPs. Gorgas et al. \shortcite{Gea97}
showed that Sphs roughly follow solar abundance  ratio SSP loci in the
$<$Fe$>$ vs Mg$_{2}$  diagnostic  plot. Low  luminosity  spiral bulges
observed by Jablonka, Martin \& Arimoto \shortcite{Ja96} also span the
region  of the Mg$_{2}$ vs $<$Fe$>$  plot occupied  by Worthey's solar
abundance ratio SSPs \cite{W98}, while  some high luminosity bulges in
their sample exhibit greater than solar [Mg/Fe].

Observations of individual stars in the bulge of our own Galaxy, while
hampered by obscuration,  have provided interesting  insights into the
population present. Minniti et   al. \shortcite{Mea95} report  a broad
range of metallicities ($-2<$[Fe/H]$<+1$)  in K giants in the Galactic
bulge    and    indicate      the   presence     of     a  metallicity
gradient. Observations of resolved M giants in the centre of the bulge
of   our own Galaxy  indicate approximately  solar abundances and also
indicate a gradient in  metallicity \cite{Frog98}. Recent episodes  of
star formation at the centre of the  bulge, with several bursts having
occurred over the  past few 100 Myr, have  also been suggested  (Blum,
Sellgren \& Depoy 1996; Frogel  1998). These observations suggest that
the population  of the  bulge of our    own Galaxy has a   complex and
extended star formation history.\\

Our long-slit observations  of    spiral bulges allow  comparison   of
central line-strengths  and gradients to those   in other galaxy types
and  constrain  possible  SFH in  the  bulges.  We   probe whether the
similarities between bulges and  elliptical galaxies in  morphological
and photometric  properties   are  the result of    similar  formation
processes, or  are  rather the  result of  the evolution of  differing
histories to  similar  present-day forms.   In Section  \ref{DR}   the
observations, data  reduction   and  error  estimates are    outlined.
Results of our  analysis  and comparison  with other Hubble  types are
reported in  section \ref{results}. Comparison  to models are detailed
in section  \ref{modelling}.  Our conclusions  and  the  direction  of
future work are outlined in section \ref{concs}.

\begin{table*}
\caption{Data  for  four spiral bulges   observed with the  Palomar 5m
telescope.  Hubble  Type and   B   magnitudes are from  de
Vaucouleurs    et al. (1991)   (hereafter  RC3). Inclinations are from
Guthrie (1992). Distances  are  from Tully (1988)  assuming H$_{o}$=75
kms$^{-1}$Mpc$^{-1}$. Recession  velocities  given are   from RC3  (HI
data) and   this work (optical  data).  The  velocity dispersions given
($\sigma_{V}$) are the  averages across  all  data points. This is   the
value used for calculation  of broadening required  to the Lick system
(see section \ref{Meas}).}
\begin{tabular}{|c|c|c|c|c|c|c|c|}
\hline
Galaxy&Hubble&Inclination&Distance&B&Recession Velocity& $\sigma_{V}$ & Scale\\
      & Type & (deg)&(Mpc)&(mag)&RC3/This work     &(kms$^{-1}$)  &(pc/arcsec)\\
      &      &           &     &     & (kms$^{-1}$)     &              &\\    

\hline
NGC 2654 &   Sab  & 90 &  23.3    & 12.8 & 1341/1320& 136  & 110    \\
NGC 3623 &   Sa   & 79 &   7.3    &  9.6 &  807/809 & 152  & 35   \\
NGC 4565 &   Sb   & 90 &   9.7    & 10.3 & 1227/1235& 155  & 46    \\
NGC 5689 &   S0/a & 78 &  35.6    & 12.7 & 2160/2221& 149  & 168    \\
\hline
\label{data}
\end{tabular}
\end{table*}

\section{Observations \& Data Reductions.}
\label{DR}

\subsection{Observations.}
\begin{figure*}
\epsfig{file=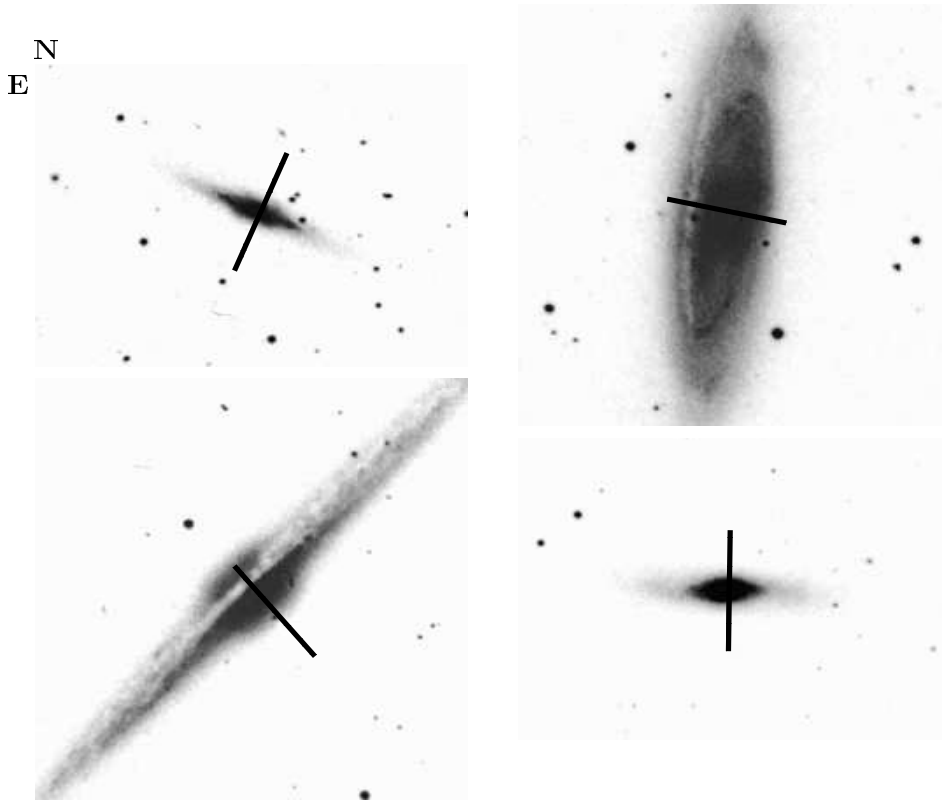, height=12cm, width=12cm, angle=0}
\caption{Digital Sky Survey images   of galaxies with  slit  positions
along minor axes. The  slit length is  2 arcmin  in all images.  Slit
position   was determined by  assuming  the  slit  passes through  the
telescope pointing position along the minor  axis. The location of the
bulge within the slit was determined by ensuring that  the peak of the
luminosity profile on   the slit matched  the position  of the maximum
count rate on  the CCD. {\bf Upper left:  NGC 2654.} Note that the slit
appears  off-centre by approximately   5   arcsec. {\bf  Upper  right:
NGC 3623.} Sky is not  reached on either side  of the  galaxy. However,
sky  was   estimated   from   extreme  edges of    slit   (see section
\ref{datred}). {\bf  Lower left: NGC 4565.}  Sky is not reached  on the
side  of the galaxy most affected  by the  disc. However, a reasonable
sky  estimate is achieved  on the  side least  affected by disc.  {\bf
Lower  right:  NGC 5689.} Note     the   slit appears  off-centre    by
approximately 5 arcsec.}
\label{slit}
\end{figure*}

Using  the Palomar 5.08  meter telescope on  1995  March 31, long-slit
spectroscopy was carried out along the minor axes of  the bulges of 4,
approximately  edge-on,  spiral galaxies (table   \ref{data} and  fig.
\ref{slit}).  We chose  to observe  edge-on  systems along their minor
axes  as disc obscuration of  bulge light is  thus concentrated on one
side of the  bulge. A  fifth  galaxy (NGC 4013)  was also observed  but
appeared to have a  bright foreground  star  slightly offset from  the
centre of the galaxy. Consequently, this galaxy was excluded from data
reductions.  The Double Spectrograph was  used \cite{OG82} with a 5700
$\AA$ dichroic filter and Texas Instruments CCDs  on both red and blue
arms.  Wavelength coverage  in the blue  was  4560 -  5430 $\AA$ at  a
resolution of 1.1 $\AA$ per  pixel.  In the  near infra-red the  range
8235 -  8860 $\AA$ was  observed at  a   resolution of 0.8  $\AA$  per
pixel. The blue wavelength range was chosen to include the H${\beta}$,
Mg   and    Fe    absorption    features   calibrated  by    Faber  et
al. \shortcite{FFBG85} and extensively observed in elliptical galaxies
\cite{DSP93,HP93,FFI96,Gea97,Vea97}.  The   red wavelength range spans
the CaII IR triplet and was intended for the study of both indices and
kinematics. Spatial resolutions of the CCDs were  0.78 and 0.58 arcsec
per pixel on  blue  and red arm  respectively.  Seeing was $\sim$  1.5
arcsec and a 1 arcsec slit was  used. Flux and velocity standard stars
were   also  observed. Object  frames  were   alternated with arc lamp
exposures to  allow for accurate wavelength  calibrations. Frames of a
continuum source (tungsten lamp) were also obtained for the purpose of
flat-fielding. Galaxy exposure times   of  5900, 7200, 4200 and   1200
seconds   were obtained  for   NGC 2654, NGC 3623,  NGC 4565 and  NGC 5689
respectively.  A maximum  single  exposure  time  of 2400  seconds was
adopted to facilitate cosmic ray removal.

\subsection{Data Reduction.}
\label{datred}
Unless   otherwise stated, data  reduction  was carried  out using the
FIGARO package of Starlink software.  A bias level from the  over-scan
region was   subtracted  from each  frame.  Due   to the  presence  of
localised features in the  flat-field frames, each object was  divided
by  the normalised, average flat-field. This  biases the data with the
spectral response of the tungsten lamp. However, this is smooth and is
removed during   flux calibration.   Cosmic   rays were   removed   by
interpolation  between adjacent   pixels.  Wavelength calibration  was
achieved by interpolating between arc lamp exposures bracketing galaxy
frames and was  accurate to $\sim$  0.1 $\AA$.  Spectra were then flux
calibrated and extinction corrected.  For NGC 2654, NGC 3623 and NGC 5689
sky   subtraction   was  carried out   by    interpolation across  the
galaxy. For NGC 3623 sky  was not fully attained  on either side of the
galaxy. This results in a small contribution  of disc light to the sky
estimate. The error associated with this sky was  estimated from the 1
sigma  variation  in index values when  sky  estimates from successive
galaxy exposures (NGC 2654, NGC 3623 and NGC 4565) were used. For NGC 4565
sky was only reached along the slit on one side of  the bulge. In this
case  the   sky   was   estimated  from    just   this  side  of   the
galaxy. Variation of the sky across the slit was shown to be less than
10\%  of  sky   level for   frames  clearly  reaching  sky    on  both
sides. Finally,  galaxy frames were  shifted  and co-added  to form  a
single frame for each galaxy.

\subsection{Measurement of Kinematics and Indices.}
\label{Meas}
Kinematics were   measured in  order   to  correct  our  data to   the
dispersion of the Lick system. Measurements of kinematics were carried
out on both red and blue data using both cross-correlation and Fourier
quotient techniques  within the IRAF  software package. Results of the
two techniques were in agreement  to within 5 kms$^{-1}$. However, the
Fourier   quotient technique was    used as   the  statistical  errors
associated with cross-correlation were substantially larger due to the
insensitivity of this method at  the low velocity dispersions  typical
of spiral bulges. As  velocity dispersions in  our sample are all less
than the  dispersion required for calibration to  the Lick system, the
additional  broadening  required  (  $\sigma_{B}$) can  be  calculated
(assuming Gaussian profiles) by;

\begin{center} 
$\sigma_{B}^2$ = $\sigma_{L}^2$ - $\sigma_{I}^2$ - $\sigma_{V}^2$\\
\end{center}

\noindent The Lick resolution ($\sigma_{L}$) is approximately constant
at $3.57   \AA$  over  the 4500-5500   $\AA$   range  covered by   our
observations \cite{WO97}.  Instrumental broadening ($\sigma_{I}$)  was
estimated   from  arc lines  and   found  to be  1.12  $\AA$. Velocity
dispersion broadening of   the galaxy ($\sigma_{V}$) was estimated  as
the average over the whole galaxy in the blue (table \ref{data}).

After broadening the spectra to the  Lick resolution, the indices were
evaluated by our own code, using the band wavelength range definitions
supplied by G. Worthey on his home  page. Worthey also kindly provides
data with   which our code could  be  tested. Differences  between the
Worthey  derived  indices and  our own for   the provided spectra were
$\leq$ 0.03 $\AA$ for  the line features  and $\leq$ 0.002 mag for the
Mg$_{1}$ and Mg$_{2}$ molecular bands. These discrepancies are smaller
than differences caused   by re-calibration of  Worthey's data  to our
wavelength resolution and  are probably the result  of the  effects of
differences   in the   handling   of  partial  bins.   They  are  also
significantly smaller than the  total errors (section \ref{error}).  A
more important source of uncertainty in our  data is the conversion to
the Lick system (which was not flux calibrated).   One of the observed
velocity standards,  HD 139669  is  also  one of the   Lick calibration
stars. Indices derived from our observations of this star and the data
of  Jones \shortcite{J96}   (data  available  on   AAS CD-ROM   Series
Vol.  VII) were  compared   to the  Gorgas  et al.   \shortcite{Gea93}
published data.  Differences between indices derived from our data and
those of Jones are typically within  errors. However, there is a large
discrepancy  between our measurement and  the  published Lick data for
the   H${\beta}$ and Fe5015  indices (also  reproduced in  the data of
Jones) which is probably caused by  unpredictable wavelength shifts in
the    Lick   IDS      spectra      (private    communication     with
J. Gorgas). Differences between our measured indices and the published
Lick  data  for this  star are given   in table \ref{errors}. HD 139669
(HR 5826) also  lies well away  from the  mean  in fig.10 of Gorgas  et
al.  \shortcite{Gea93}. Therefore,   this star   is not suitable   for
estimation   of correction factors    to   the Lick  system.  Lack  of
sufficient  data to  fully calibrate to  the Lick  system represents a
limitation on the Palomar data set. However, corrections to the narrow
line features discussed  here (Fe4668, H${\beta}$,  Fe5270 and Fe5335)
have  been shown   to   be small \cite{Gea97,Vea97,WO97}.  For  Jones'
\shortcite{J96}    observations  correction  to   the flux calibration
sensitive Mg$_{2}$ index  was found to be  small (0.01 mag increase in
the   measured  Mg$_{2}$   index). The   good   agreement between  our
measurement  of  the  Mg$_{2}$  index of  HD 139669 and   that of Jones
suggests that our  correction factor to the  Lick system should be  of
similar  magnitude ($\sim$ +0.01  mag),  although we have not  applied
this correction.

Initial measurements of  indices showed unusual gradient inversions in
the CaII, Mg$_{1}$ and  Mg$_{2}$ indices at the  centres of the bulges
(defined as  the  peak in brightness).  These   were  found to be  the
result   of     focus variation caused    by   distortion  of  the CCD
surface.   This  effect  redistributes flux  spatially   between bins,
introducing  false continuum  variations.   CaII  and molecular   band
indices are sensitive to   this type of distortion   due to the  large
separation of their sidebands. The  problem was most pronounced in the
red  (this CCD has subsequently   been  replaced at  Palomar).  As   a
result, line-strength measurements were not carried out on the red arm
data. Kinematic analysis were still possible as the continuum shape is
removed  prior to line-broadening measurement.   Stellar data from the
blue arm  had a smooth point source  FWHM  variation of $<$20\% across
the  wavelength range. Even so, features  are  still introduced in the
Mg$_{1}$  and  Mg$_{2}$ indices in  the centre  of the  galaxies where
luminosity  gradients  are greatest. A minimum  binning  of $\sim$ 1.5
arcsec  has  minimised    this    effect and    matched   the   seeing
conditions. Central values for purposes of modelling and comparison to
ellipticals  were calculated using the central   3 arcsec (see section
\ref{Lindices}).

\subsection{Estimation of Errors.}
\label{error}
\begin{table*}
\caption{Error table for NGC 3623.  {\bf Above:}  Index values for  the
central 3  arcsec and the    associated Poisson noise  are given.  Sky
errors are derived as detailed in section \ref{error}. Calculations of
both velocity dispersion and recession velocity errors ($\sigma$ and V
respectively) assume approximate   velocity uncertainties of  $\pm$ 35
kms$^{-1}$. Total errors are Poisson, sky, $\sigma$ and V errors added
in quadrature. Comparison to HD 139669 represents published Lick values
minus values obtained from our observation of this star. These offsets
are  not  included  in the calculation   of  total error due to  their
uncertainty. {\bf  Below:}  Error table  for outer regions   ($\sim$ 8
arcsec from centre) of NGC 3623.}
\label{errors} 
\begin{tabular}{|c|c|c|c|c|c|c|c|}
\hline 
\hline
\multicolumn{3}{l}{{\bf CENTRAL VALUES}}&&&&&\\
\hline
 Index      &  Index   &Poisson  &  Sky   &$\sigma$& V   &Total&Comparison\\
            &  Value   &Noise    &  Error &Error   &Error&Error&to HD 139669\\
\hline
Fe4668      & 8.64 \AA & 0.07    &   0.05 & 0.05 & 0.06 &0.12& 0.15  \\
H${\beta}$  & 1.29 \AA & 0.03    &$<$0.01 & 0.03 & 0.02 &0.05& 0.60  \\
Fe5015      & 4.91 \AA & 0.07    &   0.07 & 0.23 & 0.04 &0.25&-0.93  \\
  Mgb       & 4.51 \AA & 0.04    &   0.02 & 0.09 & 0.02 &0.10&-0.03  \\
Fe5270      & 3.31 \AA & 0.04    &   0.01 & 0.10 & 0.01 &0.11&-0.47  \\
Fe5335      & 3.07 \AA & 0.05    &$<$0.01 & 0.17 & 0.02 &0.18& 0.14  \\
Mg$_{1}$    & 0.133 mag& 0.002   &$<$0.001& 0.001&$<$ 0.001&0.002& 0.014 \\
Mg$_{2}$    & 0.279 mag& 0.002   &$<$0.001& 0.001& 0.001&0.002& 0.016 \\
\hline
\hline
\multicolumn{3}{l}{{\bf OUTER VALUES}}&&&&&\\
\hline
Fe4668      & 6.88 \AA & 0.12    &   0.28 & 0.07 & 0.05 & 0.32 & 0.15\\
H${\beta}$  & 1.22 \AA & 0.06    &   0.01 & 0.01 & 0.01 & 0.06 & 0.60\\
Fe5015      & 3.47 \AA & 0.12    &   0.47 & 0.21 & 0.11 & 0.54 & -0.93\\
  Mgb       & 4.12 \AA & 0.07    &   0.12 & 0.07 & 0.01 & 0.16 & -0.03\\
Fe5270      & 2.79 \AA & 0.08    &   0.03 & 0.09 & 0.01 & 0.12 &-0.47\\
Fe5335      & 2.59 \AA & 0.09    &   0.02 & 0.13 & 0.01 & 0.16 & 0.14\\
Mg$_{1}$    & 0.107 mag&0.003    &  0.001 & 0.001&$<$ 0.001& 0.003& 0.014\\
Mg$_{2}$    & 0.247 mag&0.004    &  0.003 & 0.001& 0.001& 0.005& 0.016\\
\hline
\hline
\end{tabular}
\end{table*}

Detailed analysis   of errors affecting  NGC 3623  are  shown  in table
\ref{errors}  for illustration. For all  galaxies, binning of data was
selected to maintain an approximately  constant Poisson error  outside
the central regions. For NGC 2654, NGC 4565 and NGC 5689, sky errors were
evaluated by varying the region of sky used for sky estimation.  The 1
$\sigma$ variation  in estimated index   value is taken as the  error.
For NGC 3623, which failed  to reach sky  on both sides of the  galaxy,
sky error  was estimated as the  1 sigma  variation in indices derived
utilising  sky estimates from  NGC 3623 and the bracketing observations
of   NGC 2654 and NGC 4565.   Variation  of both velocity dispersion and
recession velocity about the assumed average values (table \ref{data})
were both  $\sim \pm$ 35  kms$^{-1}$.  These variations  are significantly
larger than  errors    associated  with wavelength    calibration  and
statistical errors   from  the Fourier  quotient  technique  ($\sim$ 9
kms$^{-1}$).  Consequently, velocity dispersion  errors were estimated
as the variation in  indices caused by  increasing the galaxy velocity
dispersion assumed in calculation of broadening to  the Lick system by
35  kms$^{-1}$ . Index  errors  due to recession velocity  uncertainty
were  estimated as the  variation in indices  caused by increasing the
red-shift estimate by 35 kms$^{-1}$.

The magnitude of the sky error  varied between galaxies. Consequently,
unless otherwise stated, figures and  tables show combined Poisson and
sky  errors.  Velocity dispersion   and recession velocity  errors for
other galaxies were  similar in magnitude to  those in NGC 3623. In the
centres of   the bulges Poisson noise   and velocity dispersion errors
dominate.  While  for most  cases  sky error   dominates in  the outer
regions,    total  errors in  Fe5270   and   Fe5335  continue to  have
significant  contributions  from  velocity dispersion  error. Gradient
errors given in table \ref{cindices} were estimated as the statistical
errors associated with linear fitting.

\section{Results.}
\label{results}
\subsection{Kinematics.}
\label{sub}
\begin{figure*}
\epsfig{file=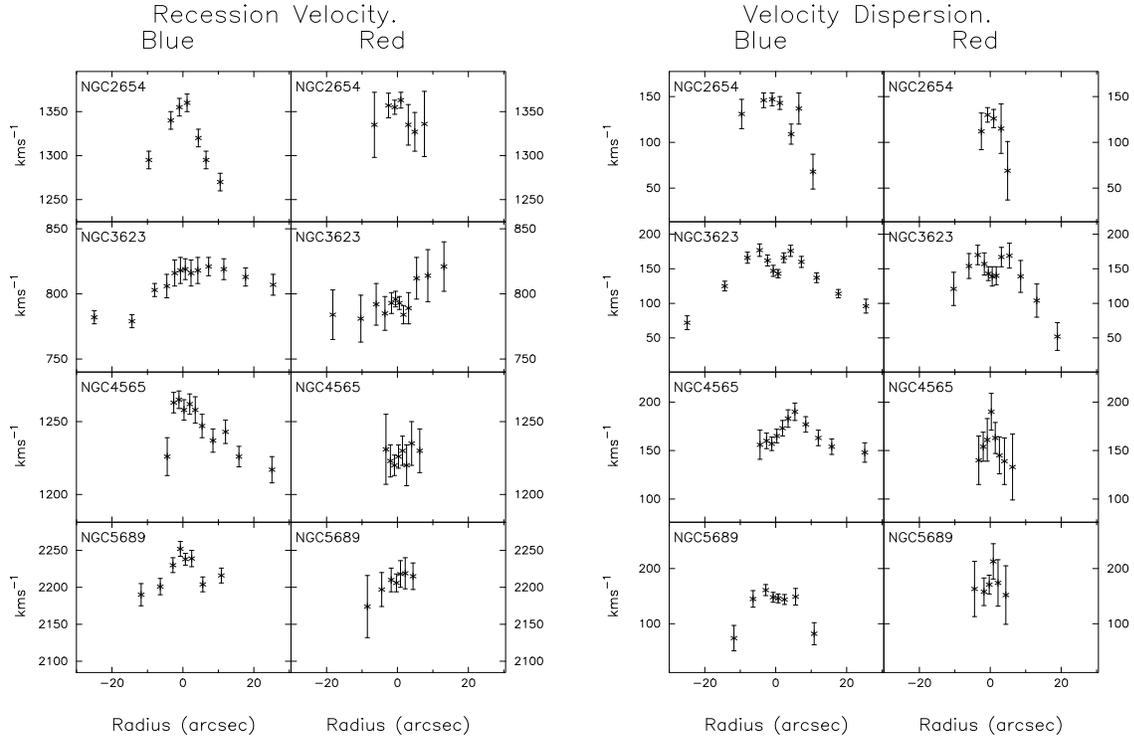, height=10cm, width=15cm, angle=0}
\caption{Spatially   resolved  kinematics along  the   minor axes of 4
spiral bulges. Recession velocities and velocity dispersions from both
the blue   and red Palomar  data  are shown. Both   sets of  data were
derived by  the   Fourier  quotient technique. Error   bars  represent
statistical  errors  of this technique and   do not include systematic
errors.}
\label{kinematics}
\end{figure*}

Results of recession  velocity  and velocity dispersion analysis   are
shown  in fig.  \ref{kinematics}.  Galaxy  centres are  defined as the
position of the luminosity peak  along the slit. Reasonable  agreement
exists between results for  red and blue data  with average values for
both recession velocity and velocity dispersion differing by less than
20  kms$^{-1}$. However,  systematic differences  are  present in  the
centres of NGC 3623 and  NGC 4565. Central recession velocities  derived
from the red data are  highly susceptible to  systematic  error due to
focus variation (section \ref{Meas}). Consequently, analysis of indices
has been based on the blue recession velocity data only. In NGC 4565 (a 
known LINER, Keel, 1983), the maximum in the blue velocity  dispersion 
profile is offset by $\sim$ 5 arcsec with respect to both the luminosity  
and  metallicity  peaks (see figs \ref{kinematics} and \ref{indices}). 
This would appear  not to  be an internal extinction effect as metallicity 
continues to rise with luminosity to the centre. Photometric  observations  
in  the near-infrared of NGC 4565 \cite{Rea96}, rather than 
indicating reddening by dust in the centre, show  a  flattening of the 
colour gradient that is  evident  along the   minor axis at  distances 
greater than 5 arcsec (Rice et al. 1996; fig. 2a). The change in colour  
gradient is coincident with the peak in  blue velocity dispersion data in
fig. \ref{kinematics} and  and the edge of the broad metallicity peak in
fig. \ref{indices}. The central region of radius 5 arcsec 
is also the emission region  of NGC 4565.  Strong [OIII] 5007 
$\AA$ and 4959 $\AA$ emission  lines are present in this region affecting
the Fe5015 index, while the H${\beta}$ index  exhibits features due to
emission line filling  (fig.  \ref{indices}).  Thus, the  data suggest
that the core of  NGC 4565 contains a  blue, metal rich, kinematically cold 
sub-population perhaps associated with ongoing star formation. The presence 
of such a population would also cause a displacement between red and blue 
luminosity peaks as observed.

A dip in velocity dispersion is seen in the centre of NGC 3623 and
is  reproduced in both red   and blue data  sets.  Such velocity
dispersion profiles have been observed along  the major axis of spiral 
galaxies (e.g. Bertola et al. 1996; Bottema \& Gerriston 1997). Kinematic
models that reproduce such dips in velocity dispersion include the  
presence  of a kinematically distinct component  within the bulge  such 
as an isothermal nuclear core \cite{BG97}, a disc \cite{Bea96} 
or a bar \cite{F96}. Chemo-dynamical modelling of the  
effect of the presence of a bar on disk galaxies \cite{F98} has shown that 
metallicity gradients are  reduced in the plane of the bar around  and 
outside the region  of the co-rotation radius, while steeper gradients 
can occur at radii  well within co-rotation. A flattening of   the 
metallicity gradients is  also   induced along the minor axis.  The 
gradients in metallicity  sensitive indices along the minor axis of 
NGC 3623 (figs \ref{indices}), which flatten rapidly away from   the  
centre,   are  consistent  with   such  models.  Friedli's \shortcite{FBK94} 
models also predict inflow of gas within the co-rotation
radius resulting in  a central burst  of metal rich star formation. The
sharp  metallicity  peak, evident  in both Fe   and Mg indices in fig
\ref{indices}, may be the result of this process. Photometric observations
by Burkhead et al. 1981 and recent Hubble archive images of NGC 3623 show 
boxy central isophotes which may also be indicative of the presence of 
a bar. Alternatively, the kinematic  feature may be  the  result of an
embedded  stellar  disc. If this is  the  case, either the disc  has a
scale-height of  $\sim$ 150 pc (major  axis disc) or the  disc rotates
along  the minor axis with a  radius of $\sim$ 150 pc. However,
there is no evidence  of the kinematic  signature of a minor axis disc
on   this    scale    in    the   recession  velocity     measurements
(fig. \ref{kinematics}).  On balance  it would  seem  likely  that the
centre of this spiral bulge harbours a small scale bar. (Note  also that 
RC3 gives the classification of NGC 3623 as SXT1. This indicates a "mixed" 
bar/no bar type).

As all the indices described here are from  the blue data, and the red
data suffer degradation  by  focus variation, we   have used the  blue
velocity dispersion and recession velocity values in our analysis of the
indices.
Average  values across the   galaxies   were used for both   recession
velocity and velocity   dispersion in  recognition of the   systematic
uncertainties.  The  resultant   recession   velocities  are in   good
agreement with values given in RC3 (table \ref{data}).\\

\subsection{Lick Indices.}
\label{Lindices}
\begin{table*}
\caption{{\bf Top:} Central  3 arcsec index  values (except H${\beta}$
for   NGC 4565 where  the  average  of outer values   is  used to avoid
emission: see section \ref{Lindices}).  Errors (given in brackets) are
Poisson noise and sky errors  combined. {\bf Middle:} Index  gradients
calculated from  all available  data points.  For  NGC 4565, errors are
statistical errors (associated with linear fitting technique) combined
with  uncertainty   in  position  of   the   galaxy centre. Positional
uncertainties are small in   other galaxies, consequently, for  these,
quoted errors are purely  statistical  errors. {\bf  Bottom:}  Central
velocity dispersions.  Quoted errors are  statistical errors  from the
Fourier quotient  technique.
$^{1}$ Peak   value 175 kms$^{-1}$ at  5 arcsec off-centre.  
$^{2}$ Peak  value 190   kms$^{-1}$  at 5  arcsec off-centre.}
\begin{tabular}{|c|c|c|c|c|}
\hline
\hline
\multicolumn{3}{l}{{\bf Central Index Values}}&&\\
\hline
                 &{\bf NGC 2654}&{\bf NGC 3623}&{\bf NGC 4565}&{\bf NGC 5689}\\
Fe4668 (\AA)     & 7.33(0.25)  & 8.64(0.08)  &8.94(0.17)   &8.72(0.26)\\
H${\beta}$ (\AA) & 1.47(0.09)  & 1.29(0.03)  &1.49(0.08)   &1.98(0.11)\\
Fe5015 (\AA)     & 4.69(0.27)  & 4.91(0.10)  &2.49(0.16)   &5.74(0.24)\\
Mgb   (\AA)      & 4.13(0.11)  & 4.51(0.04)  &4.67(0.08)   &4.03(0.13)\\
Fe5270 (\AA)     & 3.00(0.12)  & 3.31(0.04)  &3.32(0.09)   &3.00(0.14)\\
Fe5335 (\AA)     & 2.67(0.14)  & 3.07(0.05)  &3.14(0.10)   &2.62(0.16)\\
Mg$_{1}$ (mag)   & 0.105(0.005)& 0.133(0.002)&0.139(0.004 )&0.095(0.006)\\
Mg$_{2}$ (mag)   & 0.244(0.007)& 0.279(0.002)&0.293(0.005 )&0.237(0.008)\\
\hline
\hline
\multicolumn{3}{l}{{\bf Index Gradients } ($\delta$ Index/$\delta $log(radius))}&&\\
\hline
$<$Fe$>$   &-0.60(0.44)  &-0.44(0.05)  &-0.58(0.10)  &-0.55(0.19)\\
Fe4668     &-0.61(0.98)  &-1.89(0.14)  &-2.20(0.33)  &-3.28(0.60)\\
Mg$_{2}$   &-0.022(0.016)&-0.032(0.003)&-0.061(0.008)&-0.027(0.012)\\
H${\beta}$ &-0.35(0.21)  &-0.06(0.04)  & 0.52(0.11)  & 0.25(0.20)\\
\hline
\hline
\multicolumn{3}{l}{{\bf Central Velocity Dispersions}}&&\\
\hline
$\sigma_{0}$ (kms$^{-1}$)&145(7)&145$^{1}$(7)&162$^{2}$(7)&147(8)\\
\hline
\hline
\label{cindices}
\end{tabular}
\end{table*}

\begin{figure*}
\epsfig{file=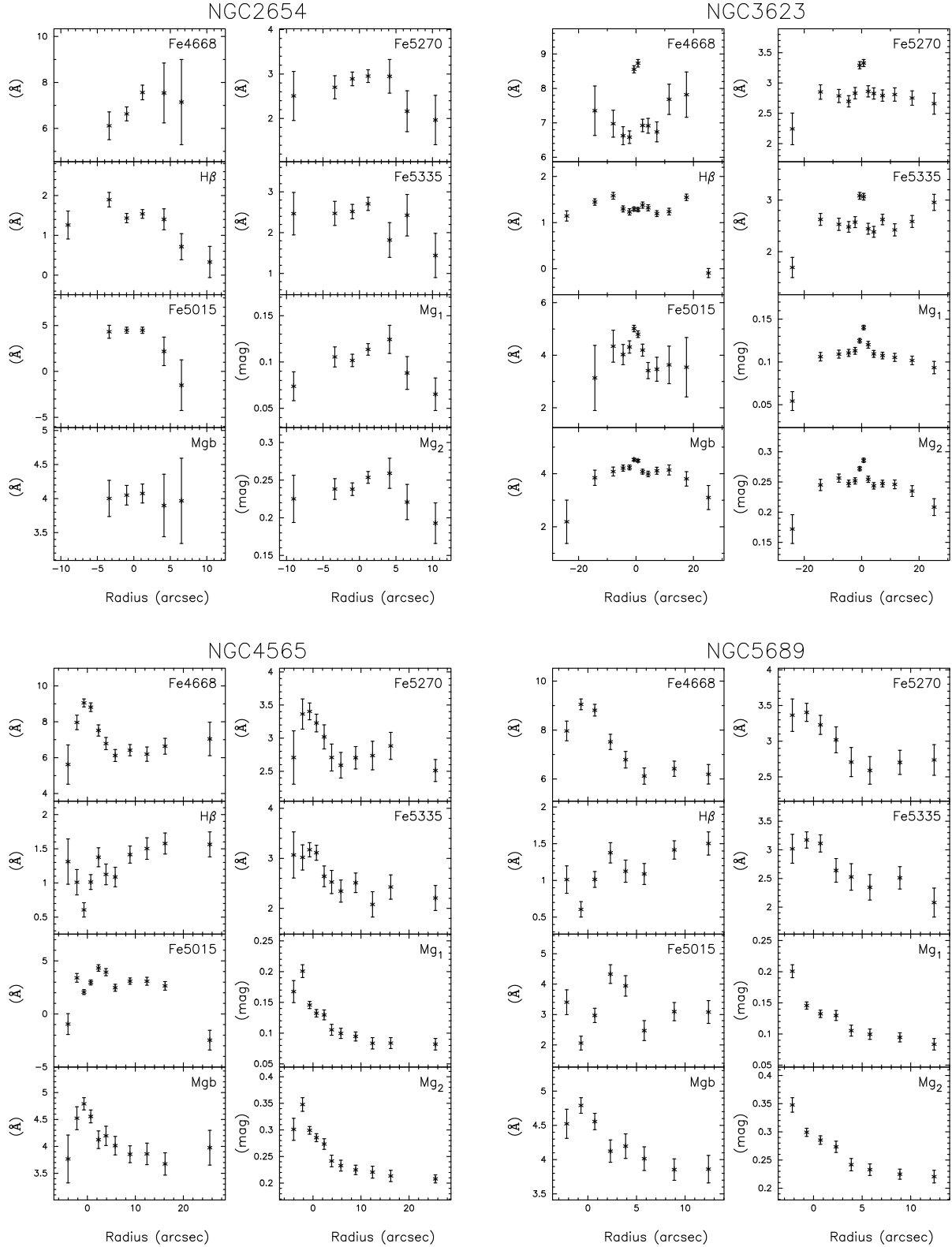, height=22cm, width=16cm, angle=0}
\caption{Indices across the  minor   axes of  four spiral  bulges.  
Radial gradients can  be  seen in  most metallicity  sensitive
features. The central dips in H${\beta}$ in  NGC 4565 are the result of
emission. Fe5015  is  also emission affected.  The  errors  bars shown
include Poisson noise and sky  error only. Bin  sizes were selected so
as to maintain approximately constant  Poisson error values beyond the
central pixels.   The increasing influence of  the  sky error at large
radii can be   seen. Note  the  sharp   central peak in    metallicity
sensitive indices   in NGC 3623.  This  peak  coincides with a  dip  in
central velocity dispersion.}
\label{indices}
\end{figure*}

In this section we  present the results  of index evaluations and plot
the data on several diagnostic plots for  comparison to models of SSPs
and observations  of other  galactic morphologies.  Index measurements
across   the   4  bulges  are  shown     in  table  \ref{cindices} and
fig. \ref{indices}. Negative radial gradients  can be seen in all  the
metallicity  sensitive indices with the  possible exception of Fe5015,
which is strongly  affected by  the  [OIII] 5007 $\AA$ emission  line.
Mgb is affected by emission of the [NI] 5199 $\AA$ doublet in one
of the side-bands \cite{GE96}. However, the  enhancement of this index
caused by emission appears to be small in these data for all galaxies,
with the possible  exception of the LINER  NGC 4565. The  age sensitive
H${\beta}$ index   can    also suffer from  emission    line  filling,
particularly near the centres of bulges. This  is most clearly seen in
NGC 4565 (fig.  \ref{indices}).  Dips  in  H${\beta}$ at  the centre of
this galaxy are associated with  two separate emission regions and are
reflected by  features in the Fe5015  index. The dip  in H${\beta}$ in
the   central   region     of   NGC 3623   may    also indicate    weak
emission.  However,   this  galaxy exhibits  sharp   central peaks  in
metallicity  sensitive features  which  coincide  with a  dip  in  the
velocity dispersion (see section \ref{sub}).   The combined effects of
the the   influence  of the metallicity  peak    and emission  on  the
H${\beta}$ index in   the centre of  this  galaxy makes interpretation
difficult. The symmetry of  indices about galaxy centres indicate that
only in NGC 4565 is disc obscuration significant.

\begin{figure*}
\epsfig{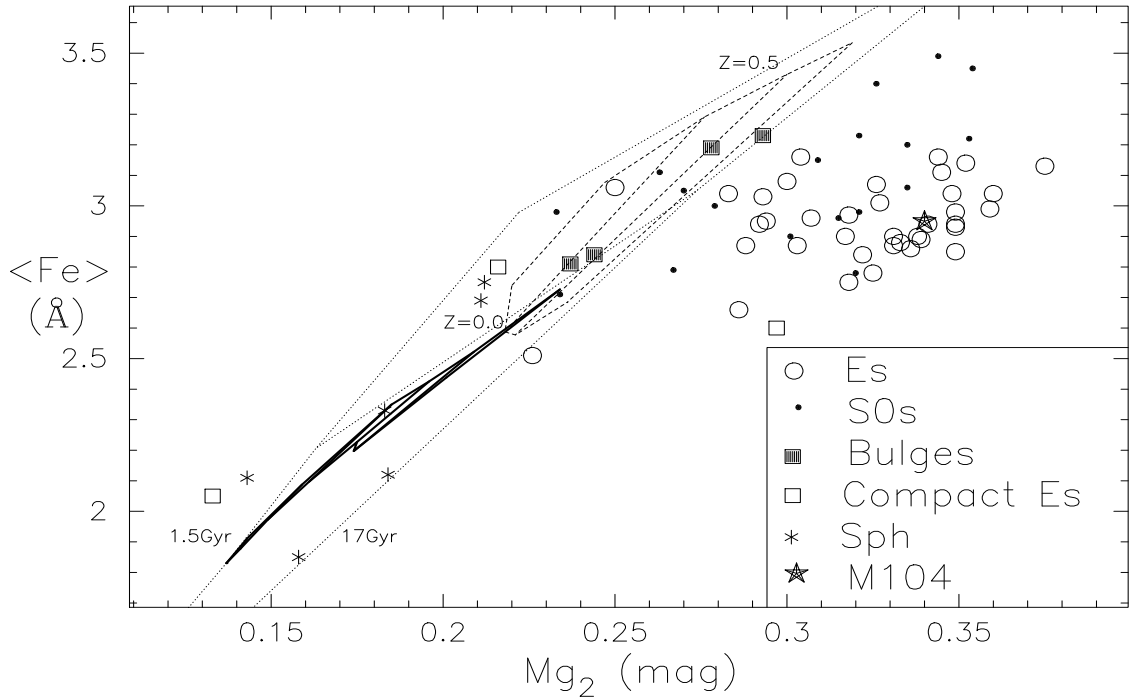}
\caption{Mg$_{2}$ vs $<$Fe$>$ for  elliptical and Sph galaxies (Gorgas
et al. 1997) and S0 galaxies (Fisher et  al. 1996).  Results for the 4
spiral bulges presented here and the bulge of  the Sa galaxy M104 (Hes
\&  Peletier   1993) are  also  shown. The  dotted  lines indicate the
extremes of Worthey's (1994) SSPs and also the solar metallicity locus
(Z=0.0). The solid  lines indicate the range  of index values achieved
by  primordial collapse models    with   a range of  star    formation
efficiencies    from 0.2 to   4.0   Gyr$^{-1}$  (described in  section
\ref{modelling}). Dashed lines indicate index values achieved
by a range of extended inflow  models with star formation efficiencies 
of 4.O Gyr$^{-1}$ over 4 to 6 Gyr. Ages of 5, 8, 12 and 17 Gyr are shown.
All models plotted assume solar 
abundance ratios.}
\label{Mg2Fe}
\end{figure*}

Fig. \ref{Mg2Fe} shows central values of  Mg$_{2}$ vs $<$Fe$>$ for our
four  spiral  bulges.   Also   plotted are central   index   values of
elliptical (E) and Sph galaxies \cite{Gea97}, S0 galaxies \cite{FFI96}
and the bulge of M104  \cite{HP93}. This diagram illustrates trends in
the relative abundance  and abundance ratio of different morphological
types.  Dotted lines  indicate the  extent  of  the space occupied  by
Worthey's solar abundance ratio SSPs and  the solar metallicity locus.
The high  central strengths of  the $<$Fe$>$ index  found in E  and SO
galaxies are also apparent   in  spiral  bulges, with bulge    indices
comparable  to    those of   SSPs     with   solar  metallicity     or
above. Fig. \ref{Mg2Fe} also demonstrates that the cores of Es exhibit
Mg$_{2}$ values in excess of those possible for SSPs with Salpeter IMF
and  solar abundance  ratios.  This is the  well   known Mg excess  in
elliptical galaxies discussed  in section \ref{intro}. S0 galaxies lie
\emph{intermediate} between Es and the SSPs.  The position of the four
spiral bulges  is consistent with  solar abundance ratios. This  is in
agreement with the result  of Sil'chenko \shortcite{S93} and Worthey's
\shortcite{W98} interpretation of the Jablonka et al. \shortcite{Ja96}
data for bulges. Outer regions of the  bulges are also consistent with
solar abundance ratio SSPs (Sansom, Proctor \& Reid 1998). Analysis of
the flux calibrated  data  of  Jones \cite{WO97} indicated    additive
corrections of +0.01 mag  and -0.06 $\AA$  for calibration of Mg$_{2}$
and   $<$Fe$>$  features  respectively  to   the   Lick system.  While
reasonable values  for this correction for  the  Palomar data may move
some of the spiral bulge data points  just outside the region of solar
ratio SSPs, the conclusion  that Mg is significantly under-abundant in
spiral bulges, when  compared to  ellipticals, remains unchanged.  The
bulge of   M104  \cite{HP93}  does  not appear    to follow  the trend
suggested by our sample (fig.  \ref{Mg2Fe}), but instead falls amongst
the elliptical galaxies. Therefore, these  data support the suggestion
by Hes and  Peletier, that the bulge   of M104 (classified as S0/a  in
RC3) more closely resembles an elliptical  galaxy than a normal spiral
bulge.\\

\begin{figure*}
\epsfig{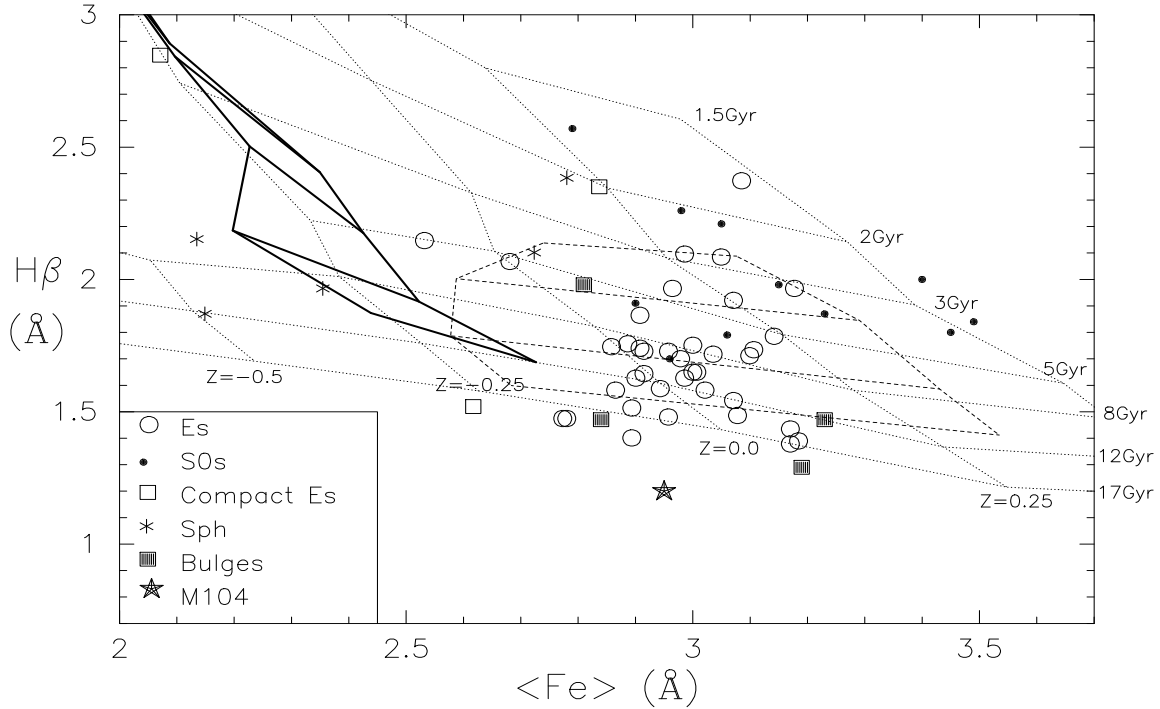}
\caption{$<$Fe$>$ vs H${\beta}$ for E, S0 and  Sph galaxies as well as
M104  and the 4 spiral  bulges presented here. Data  sources are as in
fig.\ref{Mg2Fe}. Dotted lines indicate Worthey's SSP, with values of Z
= [Fe/H]; -0.5, -0.25, 0.0, +0.25, +0.5 and ages;  1.5, 2, 3, 5, 8, 12
and 17 Gyr.  It must be  noted that due  to the effects of emission on
the  central H${\beta}$  index  for   NGC 4565 and M104,  the   plotted
H${\beta}$ values are averaged outer values. Solid and dashed lines are 
as described in fig \ref{Mg2Fe} (see section \ref{modelling}).}
\label{FeHb}
\end{figure*}

In fig.  \ref{FeHb} we plot  $<$Fe$>$ against H${\beta}$. This diagram
provides  evidence  relating    to   the relative ages   of    stellar
populations.    The correction  factors to   transform flux calibrated
H${\beta}$  data  to  the Lick  system  have been   shown to  be small
\cite{WO97}. Index values from our data are from the central 3 arcsec,
except for NGC 4565 where  the average outer  H${\beta}$ value was used
(omitting central values  where this index  is emission  affected). We
therefore     assume  no  significant   gradient    in H${\beta}$  for
NGC 4565.  This is  consistent with   studies  of elliptical   galaxies
\cite{CDB93,DSP93,FFI95,FFI96,Gea97,Vea97} that show no detectable age
gradients. The   possible presence of  emission in  the bulges creates
uncertainty in the reliability of our H${\beta}$ determination, making
interpretation of this  index  difficult. However, it should  be noted
that NGC 5689 (with H${\beta} \sim$ 2.0 $\AA$) shows no sign of central
emission which,  in any case, would  only \emph{increase} the estimate
of   H${\beta}$ for the    underlying  population.  Thus,  the  strong
H${\beta}$ in this  galaxy  is compelling evidence of  an intermediate
age ($\leq$ 5 Gyr) population.\\

\begin{figure*}
\epsfig{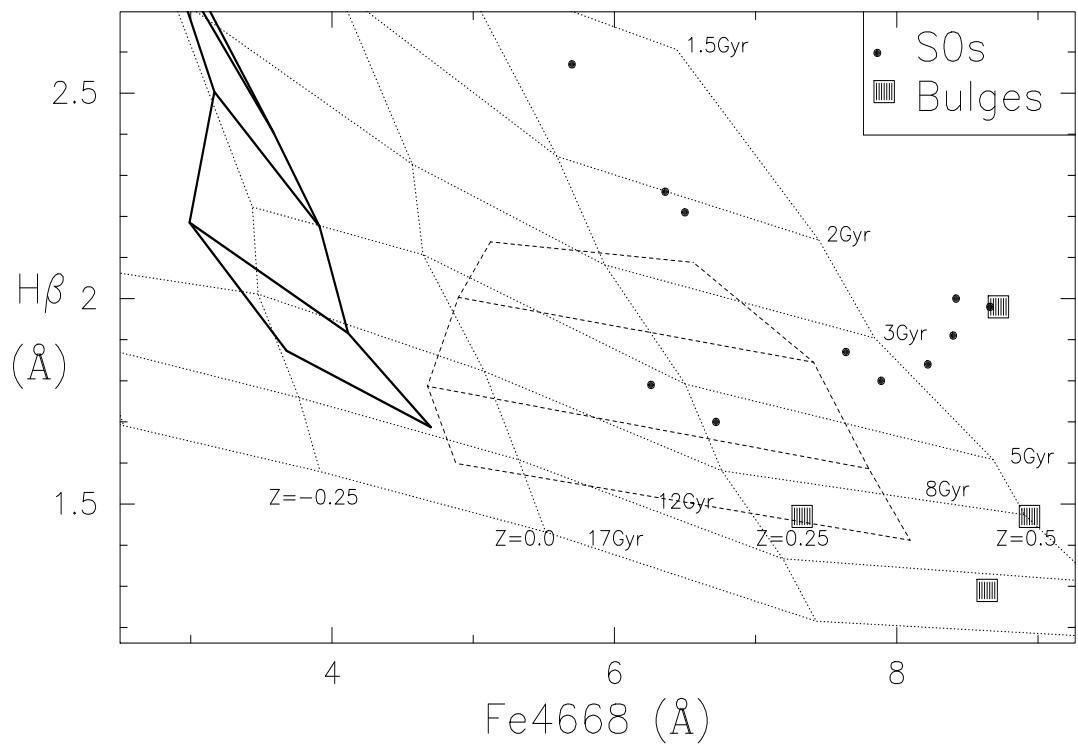}
\caption{H${\beta}$  vs Fe4668 for S0  galaxies  (Fisher et al.  1996)
and the 4 spiral  bulges presented here.  (N.B. Gorgas et  al.  (1997)
did not include Fe4668.) Dotted   lines indicate Worthey's SSP  values
for  metallicities; -0.25  0.0,  +0.25, +0.5   and ages of  1.5 to  17
Gyr. Solid and dashed lines are as in fig. \ref{Mg2Fe} (see section 
\ref{modelling}).}
\label{Fe4Hb}
\end{figure*}

Fig. \ref{Fe4Hb} locates the spiral bulges in the Fe4668 vs H${\beta}$
plane. The Fe4668 index is extremely metallicity sensitive \cite{W94}.
The correction factor to transform Fe4668 to the  Lick system has been
shown  to   be small \cite{WO97}.  Comparison of   figs \ref{FeHb} and
\ref{Fe4Hb} illustrates  that higher  metallicities  are suggested  by
Fe4668 than  by   $<$Fe$>$.  As the   Fe4668 index  is   highly carbon
sensitive \cite{TB95},  this  may indicate  either  an abundance ratio
difference or a   problem with the   calibration of Fe4668. The   main
source  of    carbon       enrichment   of the      ISM     is   still
controversial.  Gustafsson  et  al.  \shortcite{GUea99} report falling
[C/Fe] with increasing [Fe/H] in the solar neighbourhood. They suggest
that  high  mass  stars  are the  major  contributors  of  C   to  the
ISM. However,  both intermediate and low  mass stars are also known to
produce C. Consequently,  the interpretation of  a C over-abundance is
difficult. It must also be noted  that at high metallicities the value
of this index  changes rapidly with metallicity.  It  is possible that
the seeming Fe4668  excess   is in fact   the result  of the   lack of
calibrated  SSP  data  at metallicities      greater  than [Fe/H]    =
0.5. Caution  is   advised   in  interpreting  this   index   at  high
metallicities.

\subsection{Gradients in Indices and Correlations.}
\begin{figure*}
\epsfig{file=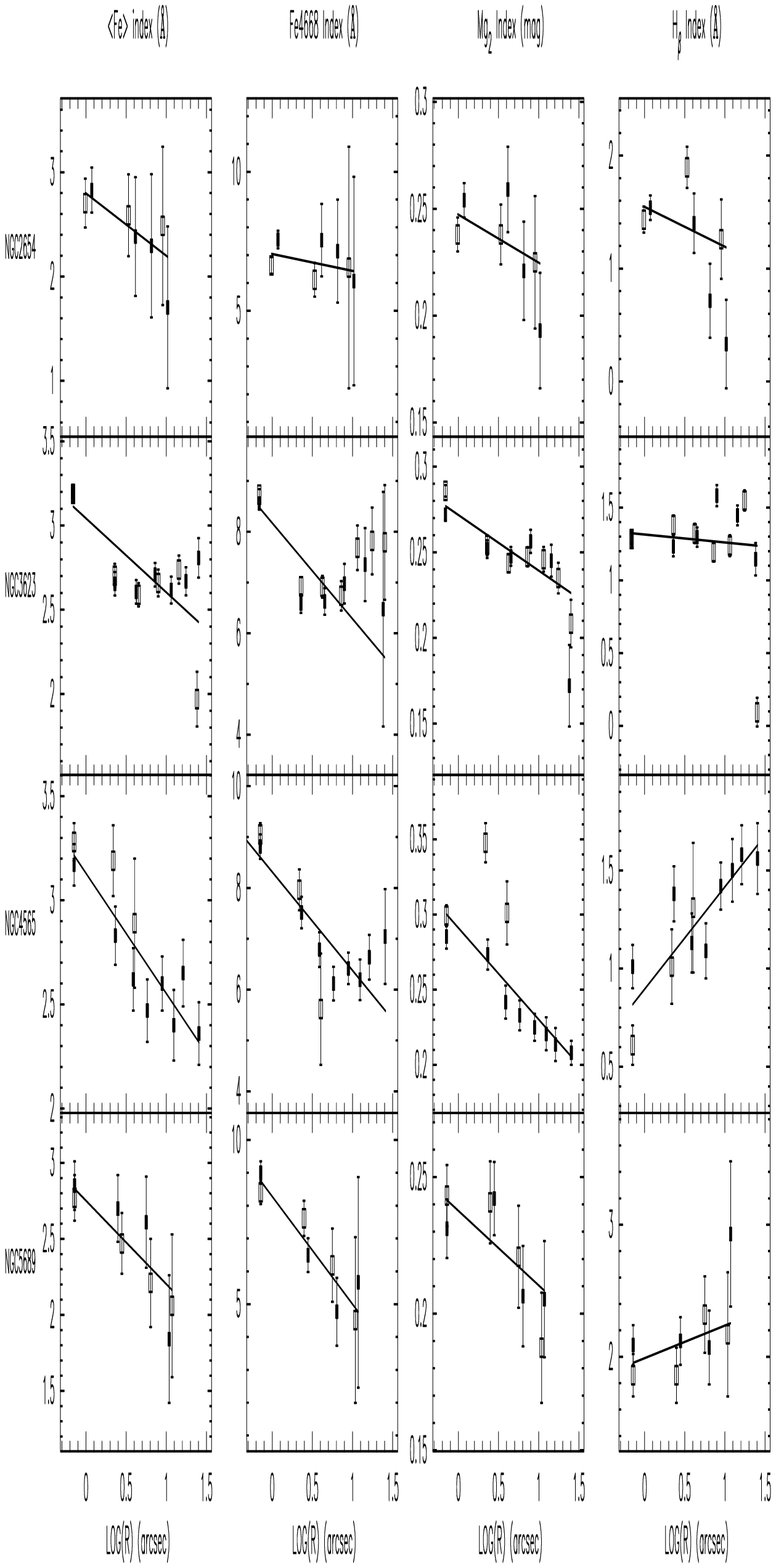, height=10cm, width=18cm, angle=0}
\caption{Index  against log radius   (radius in arcsec) for  $<$Fe$>$,
Fe4668, Mg$_{2}$  and  H${\beta}$ indices of the   4  galaxies of  the
Palomar data  set.  Open squares are  disc  contaminated  side of  the
galaxies. All data points within the regions spanned by the regression
lines are included in the   fit. Note  the flattening in   metallicity
sensitive indices in NGC 3623 which may be associated with the presence
of a bar.}
\label{grads}
\end{figure*}

Gradients were  estimated by least squares  fitting  for the $<$Fe$>$,
Fe4668, Mg$_{2}$ and H${\beta}$ indices plotted versus log radius from
the  galaxy centre  (fig. \ref{grads}). Data   from both sides of  the
galaxies were included in these fits. The gradients obtained are given
in  table   \ref{cindices}.  Central galaxy   regions  are  marginally
affected by seeing. However, central index values were included in the
gradient  estimates  due  to  the limited  number of  data  points. As
gradients are unaffected by a constant  off-set to convert to the Lick
system, a direct  comparison   with    elliptical galaxies  can     be
made. Gradients,  central   Mg$_{2}$  values  and   central   velocity
dispersions ($\sigma_{0}$;  table  \ref{cindices}) of  the bulges  are
consistent   with   the    correlations  reported   by    Carollo   et
al. \shortcite{CDB93} in elliptical galaxies. The bulges show Mg$_{2}$
gradients  of similar  magnitude   to  those in   elliptical  galaxies
possessing the same central velocity  dispersions and central Mg$_{2}$
values. The gradients are also similar  in magnitude to those reported
in  Sansom, Peace \& Dodd \shortcite{SPD94}  for 2 bulges (NGC 3190 and
NGC 1023).  The central  velocity   dispersions and the central   index
values of our Palomar  sample, indicating solar abundance ratios,  are
also consistent with the pattern  suggested by Worthey \shortcite{W98}
that \emph{all}  spheroids  with velocity  dispersions  less than  225
kms$^{-1}$ possess solar [Mg/Fe].

\section{Modelling Central Line-strengths.}
\label{modelling}

In order  to investigate possible SFHs, we  use our  galactic chemical
evolution (GCE) code detailed  in Sansom \& Proctor \shortcite{SP98a}.
Briefly,  the model calculates  star  formation  rate  (SFR) and   the
metallicity of the gas at each time-step throughout the history of the
region being modelled. As  all stars formed in each  time step are  of
the same  age and metallicity,   they   constitute an SSP.   Composite
indices are then calculated as the luminosity  weighted sum of indices
from  interpolations  between    tabulated  SSP  values   of   Worthey
\shortcite{W94} which cover populations of age 1.5 to 17 Gyr. The code 
currently models a single zone allowing for
gas in-fall.  The rate of  gas inflow can be varied  and can be either
primordial or enriched  to the  same level  as the gas  already in the
region (simulating  inflow  from  a neighbouring  region  with similar
SFH). SFR is  assumed to be proportional to  some power ($\alpha$)  of
the gas mass density ($\rho$); SFR =  C$\rho^{\alpha}$, where C is the
star   formation  efficiency  (Gyr$^{-1}$)  if   $\alpha$=1. Kennicutt
\shortcite{K89}   shows that the value of   $\alpha$ in galactic discs
lies   between  1  and 2.   In all  models   presented here  we assume
$\alpha$=1 and inflow is enriched. The differences in indices caused by 
assuming $\alpha$ = 2 were shown to be small and make no significant 
difference to the conclusions for the models presented here.
The code permits the modification of the star formation efficiency and
inflow rates at two  points in the history of  the galaxy. In this way
various    formation scenarios can    be  modelled. The star formation
efficiency modelled ranges between values for typical Sb galaxies (C =
0.2) up to  those  for star-burst  galaxy types  (C = 4).  In order to
follow the enrichment of the gas by feedback mechanisms, such as SNII,
SNIa and mass-loss  from   intermediate mass  stars, the  total  metal
content of  the gas is traced  in  a self consistent  manner. Data for
SNII (high   mass   stars)  were   taken   from Woosley    \&   Weaver
\shortcite{WW95}   and Maeder   \shortcite{M92}.  SNIa data   are from
Nomoto, Thielmann  \& Yokoi  \shortcite{NTY84}.  For intermediate mass
stars Renzini \& Voli \shortcite{RV81}  data were used.  The code also
uses the  SSPs  of Weiss,  Peletier  \& Matteucci \shortcite{Wea95} to
estimate  the effects  of non-solar abundance  ratios  on $<$Fe$>$ and
Mg$_{2}$ predictions. Currently, our GCE code  outputs estimates of 21
Lick indices for the population being modelled.

\subsection{Primordial Collapse.}
Primordial  collapse represents the   isolated collapse of a primordial
gas cloud. The  cloud commences SF when  the local density rises above
some  critical  level.  This  mechanism  has  been  proposed for   the
formation   of elliptical  galaxies.  Dynamical  models of  elliptical
galaxy  formation  by primordial collapse   \cite{C84}  show that  the
period of  gas inflow is  $<1$  Gyr. The Chemo-dynamical  models of of
Theis  et  al. \shortcite{TBH92}  also  show  the main  burst  of star
formation to be complete in $\leq$1 Gyr.  Thus, these models predict a
SFH comprising a  single burst of SF  with  a large  majority of stars
formed within the first 1 Gyr. This naturally  leads to the prediction
of [Mg/Fe] $>$ solar, observed in the  centres of elliptical galaxies,
as there is insufficient time  for SN1a to contribute large quantities
of Fe to the ISM before the bulk of SF is  complete. These models also
successfully  predict  many photometric  properties    such as the  de
Vaucouleurs r$^{1/4}$ profile and the presence of colour (metallicity)
gradients. However, several   authors  (Worthey et al. 1996;   Greggio
1997; Vazdekis et  al. 1997; Sansom  \& Proctor 1998) have  shown that
primordial   collapse  models fail  to  reproduce  the strong central,
metallicity  sensitive, line-strengths found  in elliptical  galaxies,
under the assumption  of a constant, Salpeter  IMF. This is the result
of the high number of low metallicity stars produced in this scenario.
We find  this to be  true also for the centres  of spiral bulges which
exhibit similar metallicities as   the centres of elliptical  galaxies
(fig.    \ref{Mg2Fe}). To demonstrate  this  point  we have calculated
indices for  a range of  primordial collapse models. Parameters within
the models were set  to maximise predicted central metallicities while
keeping models realistic. Thus, as in-fall  of primordial gas into the
modelled region  would dilute the  metals content of  the star forming
ISM, we assumed enriched inflow. The gas flow  rate per Gyr was set to
10   times the initial gas  content  of the  region.  As the inflow is
enriched, this large value ensures  that the number of low metallicity
stars produced is  kept small. Significantly  higher values of  inflow
rate had negligible  effects  on the  achieved line-strengths.   To be
consistent with dynamical models  of primordial collapse the period of
inflow was stopped after 0.4 Gyr while  the SF was allowed to continue
to  the  present day, albeit at   an  ever decreasing  rate  as gas is
consumed.  This generates an $\sim$  1 Gyr burst  of SF, which is also
consistent with the       chemo-dynamical  models    of   Theis     et
al.  \shortcite{TBH92}.  Continuation of  SF  to the present  day also
maximises the predicted strengths  of metallicity dependent lines.  To
demonstrate the range of  metallicities such models can  achieve, star
formation efficiency (C) during the  inflow period was varied  between
0.2 and 4.0 Gyr$^{-1}$. This range of SF efficiencies spans the values
given by Fritz-v. Alvensleben \& Gerhard \shortcite{FvAG96} for Sb, Sa
and E galaxies up  to values  for ultra-luminous star-burst  galaxies.
After the initial 0.4 Gyr of inflow, SF was allowed to continue in all
models with an efficiency C=0.2. The ages of the models varied between
1.5 and  17 Gyr. The results of  our GCE models of primordial collapse
for $<$Fe$>$, Mg$_{2}$, H${\beta}$ and Fe4668 are shown as solid lines
in figs. \ref{Mg2Fe} to  \ref{Fe4Hb}. Comparison of  indices for E, S0
and spiral bulges (fig. \ref{Mg2Fe})  to the range of values  achieved
by primordial  collapse models shows the inability  of these models to
reproduce the observed central  line-strengths  for both $<$Fe$>$  and
Mg$_{2}$ features.   This result is  confirmed  in figs \ref{FeHb} and
\ref{Fe4Hb}. Introduction  to models of  a biased IMF at  early epochs
can  resolve the under-abundance   problem but can not  simultaneously
explain  the  high  H${\beta}$   values observed  in  some  elliptical
galaxies and at least  one of our  spiral bulges (fig. \ref{FeHb}). We
might also expect to  see an Mg$_{2}$ excess with  respect to solar in
the biased IMF scenario.  There is no  evidence for this excess in our
bulge data.

\subsection{Merger Models.}
While direct evidence for primordial  collapse is not observed,
examples of  ongoing galaxy mergers are.  Indeed, Schweizer \&
Seitzer \shortcite{SS92} suggest that, based on the number of galaxies
exhibiting  on-going mergers or the    fine structure indicative of  a
recent event, $\sim$ 50\% of field ellipticals have undergone a merger
event in the last 7 Gyr. This would  seem to be  supported by the fact
that  $\sim$ 50\%  of   the   elliptical  and  S0  sample   shown   in
fig. \ref{FeHb} show H${\beta}$ values that lie \emph{above} the 8 Gyr
SSP line. In  merger models, galaxy  formation proceeds by coalescence
of fragments that have  undergone SF prior to  merger. Such models can
achieve high   central line-strengths  as  the period   of SF prior to
merger pre-enriches the  ISM from which the  bulk of the population is
formed \cite{SP98a}. It should be  noted that both primordial collapse
and merger models of galaxy  formation predict short  bursts of SF due
to    the     extremely rapid  inflow   of    gas   in  both scenarios
\cite{TBH92,BH96}.  Therefore,  the excess Mg   observed in elliptical
galaxies is  predicted by both of these  models, without the  need for
the assumption of a biased IMF. The  modelling of spiral bulges, using
merger models, is constrained by their lack of Mg$_{2}$ excess. As the
time-scale of gas  inflow  during an  interaction  is extremely rapid,
large merger induced star-bursts will result in an Mg$_{2}$ excess. As
this is not seen in our spiral bulges, if bulge formation is dominated
by  accretions/mergers,   the  SF bursts must     have been small  and
numerous. This  is consistent with the  presence of  the disc which is
highly unstable to mergers  of more  than a few  percent of  the total
galaxy mass \cite{TO92}. While   we can construct single  burst merger  models 
that reproduce the $<$Fe$>$, H${\beta}$ and Mg$_{2}$ indices observed in  
the  bulges, detailed  modelling of merger scenarios has not been carried 
out as the large parameter space implied by multiple short bursts makes these
models highly degenerate. In the limiting case, where interactions are
so  frequent that the  induced  SF  is effectively  continuous, we expect 
merger models converge to extended inflow models.\\

\subsection{Models with Extended Inflow.}

Extended inflow models represent long term inflow of gas into the region  
being modelled (e.g. Samland et al. 1997). The long duration
of the inflow is consistent with formation  of the bulge both by in-fall 
of gas from the disc (perhaps by the action of bars) as well as by the  
accretion of gas  from the halo or captured  satellite galaxies  over an 
extended period. The models assume constant SF efficiency (C) with relatively modest gas
inflow over a period of 2 to  17 Gyr and model spectral indices at a 
variety  of points in the history of the population. With reasonable 
choices for C (0.4 - 4.0 Gyr$^{-1}$)  and  inflow rate  (10$^{5}$- 10$^{6}$ 
M$_{\odot}$ Gyr$^{-1}$) the observed  $<$Fe$>$, Mg$_{2}$ and H${\beta}$ 
in  bulges can simultaneously  be reproduced  to  within errors. An inflow 
rate of 10$^{6}$ M$_{\odot}$  Gyr$^{-1}$ corresponds to an increase in mass 
in the volume modelled equal to the initial mass every Gyr. The range of SF 
efficiencies  excludes low
values (0$<$C$<$0.4) as, even with continuous SF  for 17 Gyr, there is
insufficient star formation to  produce the metals required to achieve
the  high central line-strengths   of NGC 3623 and  NGC 4565.  Formation
episodes   of $\la$ 2   Gyr result   in   an Mg$_{2}$ enhancement  not
reflected   in the  data    (using the   SSP  models  of  Weiss et al. 
1995). Figs. \ref{Mg2Fe} to \ref{Fe4Hb} show the range of indices
achieved by a selcetion of extended inflow models that reproduce the 
indices of the spiral bulges in both the Mg$_{2}$ vs $<$Fe$>$ and $<$Fe$>$ 
vs H${\beta}$ planes to within the errors. However, it should be noted 
that the models are highly  degenerate with respect  to inflow rate and 
duration, SF efficiency, and time  of SF onset (galaxy age). Consequently, 
the models are only shown to illustrate their ability to achieve the 
indices. The  range of SF  efficiencies  and inflow durations  that  can
reproduce the $<$Fe$>$,  Mg$_{2}$  and  H${\beta}$  indices in our   4
bulges are consistent  with the findings  of Samland, Hensler \& Theis
\shortcite{SHT97} using  a   chemo-dynamical  model  of    disc galaxy
formation.  Blum  et al. \shortcite{BSD96}  showed  that red giants in
the Galactic  bulge   possess  approximately solar  metallicities  and
suggest that multiple epochs of SF have  occurred in the centre of the
bulge in the last 7 - 100 Myr. Consequently, extended inflow models of
spiral  bulge formation are,   so  far, entirely consistent with  both
observations of  Lick indices,   the  stars in    our own bulge    and
chemo-dynamical modelling.

\section{Conclusions.}
\label{concs}
Our study of the bulges of 4 spiral galaxies using  the Lick system of
spectral   indices  has shown   that central  line-strengths are high.
$<$Fe$>$  indices are  similar  to those   found  in  the centres   of
elliptical  galaxies. However, a   difference in  Mg$_{2}$ index  at a
given $<$Fe$>$ exists between  elliptical  galaxies and our sample  of
spiral bulges.  Spiral bulges lie within the region of the Mg$_{2}$ vs
$<$Fe$>$ plane  occupied by the solar abundance  ratio SSPs of Worthey
\shortcite{W94},   while ellipticals  exhibit  enhanced  Mg$_{2}$ (the
known [Mg/Fe] excess).  This  difference between the two  object types
reflects differences in their star  formation histories. Our data  are
consistent with both correlations of   Mg$_{2}$ with central  velocity
dispersion  and Mg$_{2}$ gradient    with central velocity  dispersion
observed in elliptical galaxies.

Using our  GCE code, we have shown  that the central line-strengths in
spiral bulges cannot be  achieved  with primordial collapse models  of
spheroid  formation with  or  without   the assumption of    constant,
Salpeter  IMF.  Indeed, the inferred solar   [Mg/Fe]  argues against a
biased IMF in spiral bulges. Models of bulge formation with gas inflow
and star formation  extended over 2-17  Gyr can  achieve the observed
central  line-strengths  in all of    our  sample. These  models   are
consistent  with the chemo-dynamical  modelling of Samland, Hensler \&
Theis \shortcite{SHT97} as well as observations of individual stars in
the bulge of our own Galaxy \cite{Mea95,BSD96}. It has been shown that
at  least one bulge (NGC 5689) must  contain a population of relatively
young stars ($\la$ 5 Gyr) again consistent with extended inflow models
of bulge formation.

NGC 3623   shows the  presence  of  kinematic  substructure.  A  dip in
velocity  dispersion  is observed   in  the   centre  of  this  galaxy
coincident  with a sharp  peak in metallicity  sensitive indices. This
structure suggests the  presence of  a bar,  or  perhaps disc, at  the
centre of  this galaxy. Analysis of  a much larger sample of galaxies,
including  a repeat observation of  NGC 3623, is being carried out with
observations from the WHT, to test the  findings of this paper.  These
data  will be presented in  a future paper  and will include other age
sensitive indices,  such   as the  newly   calibrated H${\delta}$  and
H${\gamma}$  indices, which  are less affected  by  emission  than the
H${\beta}$ index. Increased   accuracy of  age determination,  coupled
with the larger  number of metallicity  sensitive indices, will permit
tighter constraint of  our models  of  bulge formation and their  star
formation histories. \\

{\bf Acknowledgments.}\\   We thank our referee D. Friedli for his
constructive comments. The authors acknowledge   the data analysis
facilities provide  by the Starlink  Project which is  run by CCLRC on
behalf of PPARC. In addition, the IRAF software package was used. IRAF
is distributed by the National  Optical Astronomy Observatories, which
is  operated   by AURA,  Inc.,  under cooperative   agreement with the
National  Science  Foundation. This  work  is   based on  observations
obtained at Palomar observatory,  which is owned  and operated by  the
California Institute of Technology.


\begin{thebibliography}{99}

\bibitem[\protect\citename{Balcells \& Peletier }1994]{BP94}
    Balcells M., Peletier R.F., 1994, AJ, 107,135

\bibitem[\protect\citename{Barnes \& Hernquist }1996]{BH96}
    Barnes J.E., Hernquist L., 1996, ApJ, 471,115

\bibitem[\protect\citename{Bender, Burstein \& Faber }1993]{BBF93}
    Bender R., Burstein D., Faber S.M., 1993, ApJ, 411,153

\bibitem[\protect\citename{Bertola et al. }1996]{Bea96}
    Bertola F., Cinzano P., Corsini E.M., Pizzella A., Persic M.,
    Salucci P., ApJ, 458,L67

\bibitem[\protect\citename{Blum et al. }1996]{BSD96}
    Blum R.D., Sellgren K., Depoy D.L., 1996, AJ, 112,1988

\bibitem[\protect\citename{Bottema \& Gerritsen }1997]{BG97} 
    Bottema R., Gerritsen J.P.E., 1997 MNRAS, 290,585

\bibitem[\protect\citename{Burkhead \& Hutter }1981]{BH81}
    Burkhead M.S., Hutter D.J., AJ, 86,523

\bibitem[\protect\citename{Carlberg }1984]{C84}
    Carlberg R.G., 1984, ApJ, 286,403

\bibitem[\protect\citename{Carollo et al. }1993]{CDB93}
    Carollo C.M., Danziger I.J., Buson L., 1993, MNRAS, 265,553

\bibitem[\protect\citename{Davies et al. }1993]{DSP93}
    Davies R.L., Sadler E.M. Peletier R.F, 1993, MNRAS, 
    262,650

\bibitem[\protect\citename{de Vaucouleurs et al. }1991]{dVea91}
    de Vaucouleurs G., de Vaucouleurs G., Corwin H.G., 
    Buta R.J., Fouque P., Paturel G., 1991, Third Reference 
    Catalogue of Bright Galaxies. Springer-Verlag, New York

\bibitem[\protect\citename{Faber et al. }1985]{FFBG85}
    Faber S.M., Friel E.D., Burstein D., Gaskell C.M., 
    1985, ApJS, 57,711

\bibitem[\protect\citename{Fisher et al. }1995]{FFI95}
    Fisher D., Franx M., Illingworth G., 1995, ApJ, 448,119

\bibitem[\protect\citename{Fisher et al. }1996]{FFI96}
    Fisher D., Franx M., Illingworth G., 1996, ApJ, 459,110

\bibitem[\protect\citename{Fritze-v. Alvensleben\& Gerhard }1996]{FvAG96}
    Fritze-v. Alvensleben U., Gerhard O.E., 1994, A\&A, 285,751

\bibitem[\protect\citename{Friedli et al. }1994]{FBK94}
    Friedli D., Benz W., Kennicutt R., 1994, ApJ, 430,L105

\bibitem[\protect\citename{Friedli }1996]{F96}
    Friedli D., 1996, A\&A, 312,761

\bibitem[\protect\citename{Friedli }1998]{F98}
    Friedli D., 1998, in Friedli D., Edmunds M., Robert C., Drissen L.,
    eds., ASP conf. Ser., Vol. 147, Abundance Profiles: Diagnostics Tools 
    for Galaxy History. Astron. Soc. Pac., San Francisco, p. 287

\bibitem[\protect\citename{Frogel }1998]{Frog98}
    Frogel J.A., 1998, Galaxy Evolution: Connecting the Distant
    Universe with the Local Fossil Record, Observatoire de
    Paris-Meudon, in press (astro-ph/9812034).

\bibitem[\protect\citename{Gibson \& Matteucci }1997]{GM97}
    Gibson B.K., Matteucci F., 1997, MNRAS, 291,L8

\bibitem[\protect\citename{Gorgas et al. }1993]{Gea93}
    Gorgas J., Faber S.M., Burstein D., Gonz\'{a}lez J.J., 
    Courteau S., Prosser C. 1993, ApJS, 86,153

\bibitem[\protect\citename{Gorgas et al. }1997]{Gea97}
    Gorgas J., Pedraz S., Guzm\'{a}n R., Cardiel N., 
    Gonz\'{a}lez J.J., 1997, ApJ, 481,L19

\bibitem[\protect\citename{Goudfrooij \& Emsellem }1996]{GE96}
    Goudfrooij P., Emsellem E., 1996, A\&A, 306,L45

\bibitem[\protect\citename{Greggio }1997]{G97}
    Greggio L., 1997, MNRAS, 285,151

\bibitem[\protect\citename{Gustafsson et al. }1999]{GUea99}
    Gustafsson B., Karlsson T., Olsson E., Edvardsson B., 
    Ryde N., 1999, A\&A, 342, 426

\bibitem[\protect\citename{Guthrie }1992]{G92}
    Guthrie B.N.G., 1992, A\&AS, 93,255

\bibitem[\protect\citename{Hes \& Peletier }1993]{HP93}
    Hes R., Peletier R.F., 1993, A\&A, 268,539

\bibitem[\protect\citename{Jablonka et al. }1996]{Ja96}
    Jablonka P., Martin P., Arimoto N., 1996, AJ, 112,1415

\bibitem[\protect\citename{Jones }1996]{J96}
    Jones L., 1996,  PhD thesis, Univ. North Carolina

\bibitem[\protect\citename{Keel }1983]{K83}
    Keel W.C., 1983, ApJ, 466,486

\bibitem[\protect\citename{Kennicutt }1989]{K89}
    Kennicutt R.C., 1989, ApJ, 344,685

\bibitem[\protect\citename{Maeder }1992]{M92}
    Maeder A., 1992, A\&A, 264,105

\bibitem[\protect\citename{Minniti et al. }1992]{Mea95}
    Minniti D., Olszewski E., Liebert J., White S.D.M., Hill J.M., Irwin
    M.J., 1995, MNRAS, 277,1293

\bibitem[\protect\citename{Nomoto et al. }1984]{NTY84}
    Nomoto K., Thielmann F.K., Yokoi K., 1984, ApJ, 286,644

\bibitem[\protect\citename{Oke \& Gunn }1982]{OG82}
    Oke J.B., Gunn J.E., 1982, PASP, 94,586

\bibitem[\protect\citename{Renzini \& Voli }1981]{RV81}
    Renzini A., Voli M., 1981, A\&A, 94,175

\bibitem[\protect\citename{Rice et al. }1996]{Rea96}
    Rice W., Merrill K.M., Gatley I., Gillett F.C., 1996 AJ, 112,114
    
\bibitem[\protect\citename{Samland et al. }1997]{SHT97}
    Samland M., Hensler G., Theis Ch., 1997, ApJ, 476,544

\bibitem[\protect\citename{Sansom, Peace \& Dodd }1994]{SPD94}
    Sansom A.E., Peace G., Dodd M., 1994, MNRAS, 271,39

\bibitem[\protect\citename{Sansom \& Proctor }1998]{SP98a}
    Sansom A.E., Proctor R.N., 1998, MNRAS, 297,953

\bibitem[\protect\citename{Sansom et al. }1998]{SP98b}
    Sansom A.E., Proctor R.N., Reid I.N. 1998, in Friedli D., Edmunds M.,
    Robert C., Drissen L., eds., ASP conf. Ser., Vol. 147, Abundance 
    Profiles: Diagnostics Tools for Galaxy History. Astron. Soc. Pac., 
    San Francisco, p. 26

\bibitem[\protect\citename{Sil'chenko }1993]{S93}
    Sil'chenko O.K., 1993, Pis'ma Astrom. Zh, 19,701

\bibitem[\protect\citename{Schweizer \& Seitzer }1992]{SS92}
    Schweizer F., Seitzer P., 1992, AJ, 104,1039

\bibitem[\protect\citename{Theis et al. }1992]{TBH92}
    Theis C., Burkert A., \& Hensler G., 1992, A\&A, 265,465

\bibitem[\protect\citename{T\'{o}th \& Ostriker }1992]{TO92}
    T\'{o}th G., Ostriker J.P., 1992, A\&A, 265,465

\bibitem[\protect\citename{Tripicco \& Bell }1995]{TB95}
    Tripicco M.J., Bell R.A., 1995, AJ, 110,3035

\bibitem[\protect\citename{Tully }1988]{T88}
    Tully R.B., 1988, Nearby Galaxies Catalogue, Camb. Univ. Press,
    Cambridge

\bibitem[\protect\citename{Vazdekis et al. }1997]{Vea97}
    Vazdekis A., Peletier R.F., Beckman J.E., Casuso E., 1997, 
    ApJS, 111,203

\bibitem[\protect\citename{Weiss et al. }1995]{Wea95}
    Weiss A., Peletier R.F., Matteucci F., 1995, A\&A, 296,73

\bibitem[\protect\citename{Woosley \& Weaver }1995]{WW95}
    Woosley S.E., Weaver T.A., 1995, ApJS, 101,181

\bibitem[\protect\citename{Worthey }1994]{W94}
    Worthey G., 1994, ApJS, 95,107

\bibitem[\protect\citename{Worthey }1998]{W98}
    Worthey G., 1998, PASP, 110,888

\bibitem[\protect\citename{Worthey, Dorman \& Jones }1996]{WDJ96}
    Worthey G., Dorman B., Jones L.A., 1996, AJ, 112,948

\bibitem[\protect\citename{Worthey, Faber \& Gonz\'{a}lez }1992]{WFG92}
    Worthey G., Faber S.M., Gonz\'{a}lez, J.J., 1992, ApJ, 398,69

\bibitem[\protect\citename{Worthey et al. }1994]{Wea94}
    Worthey G., Faber S.M., Gonz\'{a}lez, J.J., Burnstein D., 1994, ApJS, 94,687

\bibitem[\protect\citename{Worthey \& Ottaviani }1997]{WO97}
    Worthey G., Ottaviani D.L., 1997, ApJS, 111,377

\label{lastpage}
\end{thebibliography}
\end{document}